\documentclass[%
twocolumn,
amsmath,amssymb,
aps,
pre
]{revtex4-2}

\usepackage{graphicx,xcolor}
\usepackage{dcolumn}
\usepackage{bm}
\usepackage{amsmath,amsfonts,amssymb}

\usepackage{tikz}
\usepackage{array}
\usepackage{hhline}
\usepackage{adjustbox}
\usepackage{footmisc}
\usepackage{tabularx}
\usetikzlibrary{positioning}

\usepackage{blindtext}

\begin{document}

\preprint{APS/123-QED}

\title{The configurational entropy of random trees}

\author{Pieter H. W. van der Hoek}
\email{pvanderh@sissa.it}
\affiliation{%
SISSA - Scuola Internazionale Superiore di Studi Avanzati, Via Bonomea 265, 34136 Trieste, Italy
}%

\author{Ralf Everaers}
\email{ralf.everaers@ens-lyon.fr}
\affiliation{
ENS de Lyon, CNRS, Laboratoire de Physique (LPENSL UMR5672) et Centre Blaise Pascal, 69342 Lyon cedex 07, France
}%

\author{Angelo Rosa}
\email{anrosa@sissa.it}
\affiliation{%
SISSA - Scuola Internazionale Superiore di Studi Avanzati, Via Bonomea 265, 34136 Trieste, Italy
}%

\date{\today}

\begin{abstract}
We present a graph theoretical approach to the configurational statistics of random tree-like objects, such as randomly branching polymers.
In particular, 
for ideal trees 
we show that Pr\"ufer labelling provides:
(i) direct access to the exact configurational entropy as a function of the tree composition,
(ii) computable exact expressions for partition functions and important experimental observables for tree ensembles with controlled branching activity
and
(iii) an efficient sampling scheme for corresponding tree configurations and arbitrary static properties. 
\end{abstract}

\maketitle


{\it Introduction} --
Random tree-like structures are ubiquitous in Nature~\cite{FisherPRL1978,KurtzeFisherPRB1979,IsaacsonLubensky,DuarteRuskin1981,BovierFroelichGlaus1984,Burchard1999,RubinsteinColby} and have been intensely studied  in analytical~
{\cite{ZimmStockmayer49,DeGennes1968,LubenskyIsaacson1979,Lubensky1981,DaoudJoanny1981,ParisiSourlasPRL1981,Stilck1992,GujratiJCP1993,GujratiPRL1995,BanchioSerra1995,GrosbergNechaev2015} 
as well as computational works~\cite{Redner1979,SeitzKlein1981,Caracciolo1985,MadrasJPhysA1992,GrassbergerJPhysA2005,Rosa2016a,Rosa2016b,Rosa2016c,Ghobadpour2021,Amoebapaper2024}.
The most obvious example from soft matter are randomly branching polymers, but annealed~\cite{GutinGrosberg93,EveraersGrosbergRubinsteinRosa} tree-like structures also emerge in dense solutions of non-concatenated and unknotted circular (ring) polymers~\cite{KhokhlovNechaev85,RubinsteinPRL1986,RubinsteinPRL1994,GrosbergSoftMatter2014,RosaEveraersPRL2014,MichielettoSoftMatter2016,RosaEveraers2019,SchramSM2019,SmrekRosa2019}.
In a biological context, they represent generic models for describing the large-scale conformational properties of 
bacterial DNA~\cite{MarkoSiggia1994,MarkoSiggiaSuperCoiledDNA1995,odijk_osmotic_1998,Cunha2001,Junier2023},
chromosomal DNA during interphase~\cite{grosbergEPL1993,RosaPLOS2008,Vettorel2009,halverson2014melt}
and viral RNA~\cite{GelbartPNAS2008,Fang2011,Vaupotic2023}.

Here we are concerned with the most fundamental quantity, the Boltzmann configurational entropy,
\begin{equation}\label{eq:BoltzmannEntropy}
S = k_B \log(\Omega) \, ,
\end{equation}
of non-interacting ``ideal'' trees, where $k_B$ is the Boltzmann constant. 
The question of how many tree connectivities exist goes at least back to 1889 and Cayley, who conjectured that there are
\begin{equation}\label{eq:Cayley}
\Omega_N = N^{N-2} 
\end{equation}
distinct labelled trees of $N$ nodes~\cite{Cayley}. 
Expanding on an earlier attempt to estimate tree sizes using Pr\"ufer codes~\cite{PrueferAlgoRNA}, we revisit Pr\"ufer's proof of Cayley's conjecture to extend his reasoning, which is based on a clever one-to-one correspondence between trees and sequences (or, ``codes'') of node labels defining the chain connectivity.
In particular, we derive exact expressions for the multiplicity $\Omega$ for trees as a function of their composition as well as for important experimental observables for tree ensembles with controlled branching activity.
In addition we present an efficient sampling scheme for corresponding tree configurations which enables the study of arbitrary static properties close to the asymptotic limit.

{\it Tree configurations and spatial conformations} --
The {\it configuration} or connectivity $\mathcal G$ of a given labelled tree can be represented by a disordered double list of the $N-1$ bonds specifying the index pairs $(i,j)$ of the connected nodes $i\ne j\in[1, \dots, N]$.
Denoting with $f_i$ the number of bonds emanating from individual nodes $i$ (the so called {\it functionality} or {\it vertex degree}) and with $N_f$ the total number of nodes of functionality $f$, the latter have to add up to the total number of nodes $N$,
\begin{equation}\label{eq:Sumnf_unres}
\sum_{i=1}^N 1 = \sum_{f=1}^{N-1} N_f = N \, ,
\end{equation}
while the former sum to twice the total number of bonds,
\begin{equation}\label{eq:SumFi}
\sum_{i=1}^N f_i = \sum_{f=1}^{N-1} f \, N_f = 2(N-1) \, .
\end{equation}
In turn, Eqs.~\eqref{eq:Sumnf_unres} and~\eqref{eq:SumFi} imply that the total number of endpoints (or ``leaves''), $N_1$, satisfies the relation
\begin{equation}\label{eq:Endpoints_unres}
N_1 = 2 + \sum_{f=3}^{N-1} (f-2) N_f \, .
\end{equation}
A complementary set $\Gamma = \{ \vec r_{i=1, \dots, N} \}$ of spatial coordinates of the individual nodes defines the {\it conformation} of the polymer in the embedding space.
In particular, the latter can be chosen as a common regular lattice (simple cubic, face-centered cubic, etc...).

{\it Pr\"ufer codes} --
In 
1918 
Pr\"ufer~\cite{Prufer} found a systematic way to order the bond list (and, hence, to number the bonds) by requiring that $j$ always be chosen as the ``leaf'' with the lowest label among the remaining nodes for which all but one bond have already been specified.
The Pr\"ufer code only retains the indices $i$ of the nodes to which the leaves are connected, since the latter's identity is implicitly defined by the encoding and decoding procedures (for details and examples of these two procedures, see Sec.~S1A and Sec.~S1B 
in Supplemental Material (SM~\cite{SMnote})).
A given node $i$ appears $f_i-1$ times in the Pr\"ufer code. 
The total length of the code is $N-2$, because there is no freedom in the choice of the last bond joining the two leaves remaining after all but one of the $N-1$ bonds are specified.

{\it Pr\"ufer counting} --
The one-to-one correspondence between Pr\"ufer codes and labelled trees has the important consequence that, by counting codes, one may count trees.
In particular~\cite{Prufer}, Eq.~\eqref{eq:Cayley} follows from the fact that each of the $N-2$ elements of the Pr\"ufer code can be freely chosen among the indices of the $N$ nodes constituting the tree. 
To proceed, first consider nodes with fixed labels and functionalities, $\left\{f_i\right\}$.
They can be combined into~
\cite{Moon_1964}: 
\begin{equation}\label{eq:Omega_fi}
\Omega_N(\left\{f_i\right\}) = (N-2)! \prod_{i=1}^N  \frac{1}{(f_i-1)!} 
\end{equation}
distinct labelled trees, because there are $(N-2)!$ different permutations for the Pr\"ufer code, but permutations of identical indices do not alter the connectivity $\mathcal G$~\footnote{
As a check, one recovers Eq.~\eqref{eq:Cayley} by summing over all possible combinations of node functionalities,
\begin{equation*}
\Omega_N = \sum_{\protect\substack{f_1+f_2+...+f_N=2(N-1) \\ f_1, \, f_2, \, \cdots, \, f_N\geq 1}} \, \Omega_N(\left\{ f_i \right\}) \, ,
\end{equation*}
by setting $k_i=f_i-1$ in the multinomial theorem (en.wikipedia.org/wiki/Multinomial\_theorem) 
which states that for any integer $N\geq 2$,
\begin{equation*}
N^{N-2} = \sum_{\protect\substack{k_1+k_2+...+k_N=N-2 \\ k_1, k_2, ..., k_N\geq 0}} \, \frac{(N-2)!}{k_1! \, k_2! \dots k_N!} \, .
\end{equation*}
}.
Next, consider permutations of the tree labels.
They preserve the tree composition, $\{ N_f \}$, and lead to a different set of labelled trees {\it if and only if} they permute labels of nodes of {\it different} functionalities.
Grouping together all 
$ N! \prod_{f=1}^{N-1} \frac{1}{N_f!} $
distinct possibilities of assigning labels, one obtains the number~
\cite{Riordan1966} 
\begin{equation}\label{eq:Multiplicity_unres}
\Omega_N(\{ N_f \} ) = N! (N-2)! \, \prod_{f=1}^{N-1} \frac{1}{N_f! \, ((f-1)!)^{N_f}}
\end{equation}
of labelled trees for a given tree composition, $\{ N_f \} $.
In the rest of this work, for simplicity and with no substantial loss of generality~\cite{Rosa2016a,Rosa2016b,Rosa2016c,Ghobadpour2021,Amoebapaper2024}, we restrict the functionality of branch-nodes to $f \leq 3$, in which case Eq.~\eqref{eq:Multiplicity_unres} reduces to
\begin{equation}\label{eq:OmegaNn3}
\Omega_N(N_3) = \frac{N!}{N_1! \, N_2! \, N_3!} \frac{(N-2)!}{2^{N_3}} \, ,
\end{equation}
with the relations (see Eqs.~\eqref{eq:Sumnf_unres}-\eqref{eq:Endpoints_unres}): $N_1 = N_3+2$, $N_2 = N - 2N_3 - 2$ and maximum number of branch-nodes $N_{3, {\rm max}}(N) = (N-2)/2$ or $N_{3, {\rm max}}(N) = (N-3)/2$ for, respectively, $N$ even or odd.
For future reference, note also that according to Eq.~(\ref{eq:OmegaNn3}) with $N_1=2$, $N_2=N-2$ and $N_3=0$ there are
\begin{equation}\label{eq:OmegaLinearChains}
\Omega_N(N_3=0) = \frac{N!}2
\end{equation}
distinct labelings for a linear chain: any labelling can be read forwards and backwards because of the impossibility to distinguish the two ends.

{\it Grand canonical ensemble for branch-nodes} --
For most applications we are more interested in ensembles where the number of branch-nodes is controlled by a suitable ``chemical'' potential $\mu$ in the Hamiltonian ${\mathcal H} = -\mu N_3$~\cite{Rosa2016a,Rosa2016b,Rosa2016c,Amoebapaper2024}.
With $\beta^{-1} = k_BT$ where $T$ is the temperature and Eq.~\eqref{eq:OmegaNn3}, the partition function reads~
\cite{GujratiJCP1993} 
\begin{equation}\label{eq:BranchingPartitionFunction}
Z_{N, \mu} \equiv \sum_{N_3=0}^{N_{3,\rm max}} \Omega_N(N_3) \, e^{\beta\mu N_3} \, ,
\end{equation}
so that the probability to observe a particular number of branch-nodes is given by
\begin{equation}\label{eq:p(n_3)}
p_{N, \mu}(N_3) = \frac{\Omega_N(N_3) \, e^{\beta\mu N_3} }{Z_{N, \mu}} \ .
\end{equation}
\begin{figure}
\includegraphics[width=0.50\textwidth]{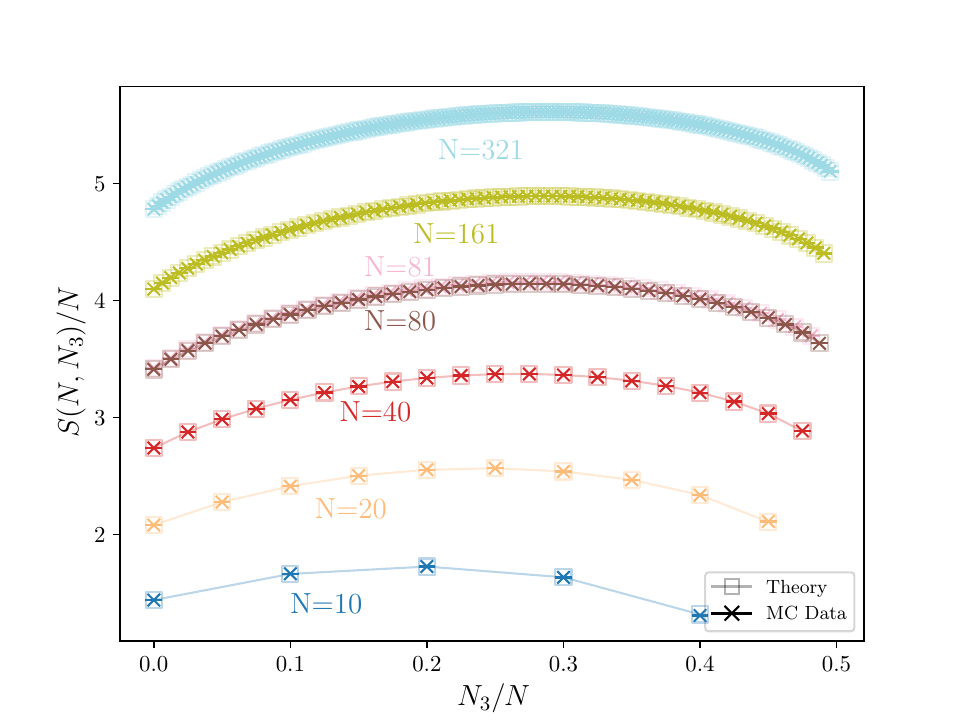}
\caption{
Connectivity entropy per node, $S_N(N_3)/N$, as a function of the fraction of branch-nodes, $N_3/N$.
Comparison between numerical results from thermodynamic integration of Monte Carlo simulations of lattice trees on the FCC lattice ($\times$, error bars are smaller than symbols' size) and Eq.~\eqref{eq:BoltzmannEntropy} with Eq.~\eqref{eq:OmegaNn3} ($\square$).
Lines serve as a guide for the eye. 
}
\label{fig:Entropies}
\end{figure}
%

{\it Validation and entropy of random trees} --
We have recently~\cite{Amoebapaper2024} sampled the above ensemble in Monte Carlo (MC) simulations of a generic lattice model for randomly branching polymers on the $3d$ FCC lattice for parameter combinations covering the complete crossover from the linear chain limit with $\langle N_3\rangle \ll 1$ to the high-branching regime with $\langle N_2\rangle \ll 1$ for chains of length up to $N=321$.
Binning our old data into normalised histograms $h_{N, \mu}(N_3)$ for the available combinations of $N$ and $\beta\mu$ and building on Eq.~\eqref{eq:p(n_3)}, we can estimate ratios of tree multiplicities, 
\begin{equation}\label{eq:NonZeroHeightBins}
\frac{\Omega_N(N_3)}{\Omega_N(N_3+1)}  \simeq \frac{h_{N, \mu}(N_3)}{h_{N, \mu}(N_3+1)} e^{\beta\mu}  \, ,
\end{equation}
and corresponding differences of the configurational entropy,
\begin{multline}\label{eq:Relative_entropies}
\lefteqn{ \frac{S_N(N_3 +1) - S_N(N_3)}{k_B} \simeq } \\
-\beta\mu + \log(h_{N, \mu}(N_3+1)) - \log(h_{N, \mu}(N_3)) \, .
\end{multline}
%
Taking a weighted average over independent estimates of $S_N(N_3 +1) - S_N(N_3)$, we can then (thermodynamically) ``integrate'' \cite{FrenkelSmit2002} the configurational entropy for all values of $0 < N_3 \le N_{3, {\rm max}}(N)$ starting from the initial value $S(N, N_3=0) = k_B \log(N!/2)$ for linear chains.
The results for different $N$ show perfect overlap with the analytical expression $S_N(N_3) = k_B \log(\Omega_N(N_3))$ (Fig.~\ref{fig:Entropies}, ``$\times$'' (simulations) {\it vs.} ``$\square$'' (theory)).

\begin{figure}
\includegraphics[trim={0.2cm 1.5cm 0 1.3cm},clip,width=0.48\textwidth]{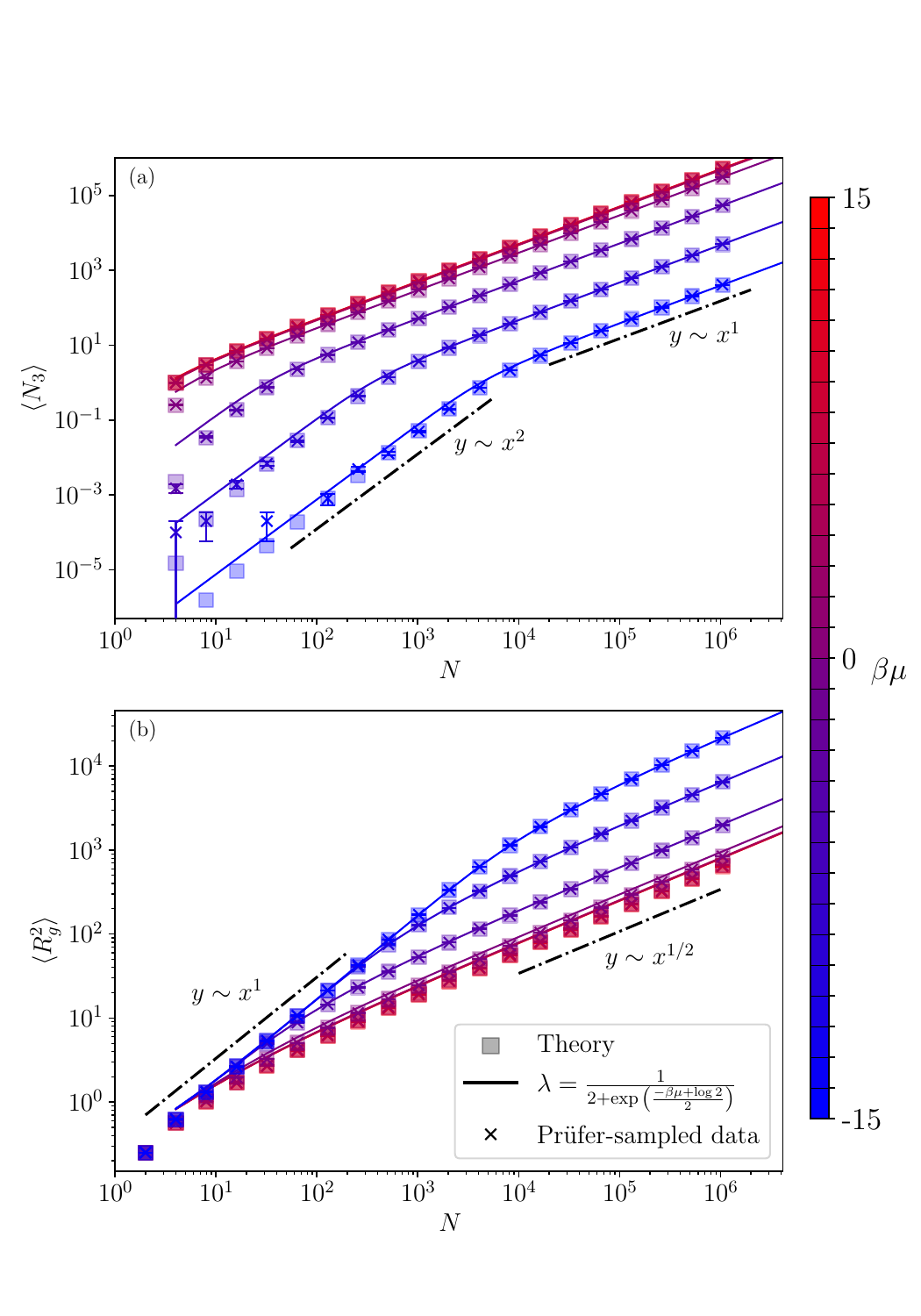}
\caption{
(a) mean number of branch-nodes, $\langle N_3\rangle$,
and
(b) mean-square gyration radius, $\langle R_g^2\rangle$, as a function of tree nodes $N$.
Results for:
($\blacksquare$) exact numerical evaluation,
($\times$) Pr\"ufer-sampled trees,
(lines) predictions of the Daoud-Joanny theory~\cite{DaoudJoanny1981} with branching activity $\lambda^2 = (2+\exp(-(\beta\mu-\log(2))/2))^{-2}$ (Eq.~\eqref{eq:LambdaFromMu}).
Different colors correspond to different values of the branching chemical potential $\mu$ (colorcode on the right).
}
\label{fig:n3+Rg2_usVSDJ}
\end{figure}
%

{\it Tree observables} -- 
Given Eqs.~\eqref{eq:OmegaNn3} and~\eqref{eq:BranchingPartitionFunction}, it is straightforward (Sec.~S2 
in SM~\cite{SMnote}) to numerically calculate the mean number of branch-nodes, $\langle N_3\rangle$, while the mean square gyration radius, $\langle R_g^2 \rangle$, follows from an extension of Kramers' theorem for {\it rooted} trees (Sec.~S3 
in SM~\cite{SMnote}). 
In panels (a) and (b) of Fig.~\ref{fig:n3+Rg2_usVSDJ} we report results for $\langle n_3\rangle$ and $\langle R_g^2\rangle$ as a function of $N$ and different $\beta\mu$ from numerical calculations ($\blacksquare$) and from Pr\"ufer-sampled trees ($\times$, 
see below for details on the procedure to generate trees by Pr\"ufer sampling and their embedding on the $3d$ FCC lattice
), the two datasets agree within the statistical error (Fig.~S1 
in SM~\cite{SMnote}).
For comparison, we show also results (lines) obtained by using the field-theoretic partition function by de Gennes~\cite{DeGennes1968} and Daoud and Joanny~\cite{DaoudJoanny1981} (reviewed in Sec.~S4 
in SM~\cite{SMnote}).
The theory introduces a single phenomenological parameter for a branching activity, $\lambda^2$, which we have chosen as
\begin{equation}\label{eq:LambdaFromMu}
\lambda = (2+\exp(-(\beta\mu-\log(2))/2))^{-1} \, ,
\end{equation}
to asymptotically match the number of branch points in our systems (Sec.~S4 
in SM~\cite{SMnote}).
The agreement is qualitative for highly branching trees and quantitative for weakly branching trees,  where the employed continuum approximation directly applies.
Notably, the data reproduce the predicted~\cite{DaoudJoanny1981,Rosa2016a} characteristic behaviors: (i) $\langle N_3 \rangle \sim N^2$ and $\langle R_g^2 \rangle \sim N$ in the low-branching regime and (ii) $\langle N_3 \rangle \sim N$ and $\langle R_g^2 \rangle \sim N^{1/2}$ in the high-branching regime.

\begin{figure}
\includegraphics[width=0.50\textwidth]{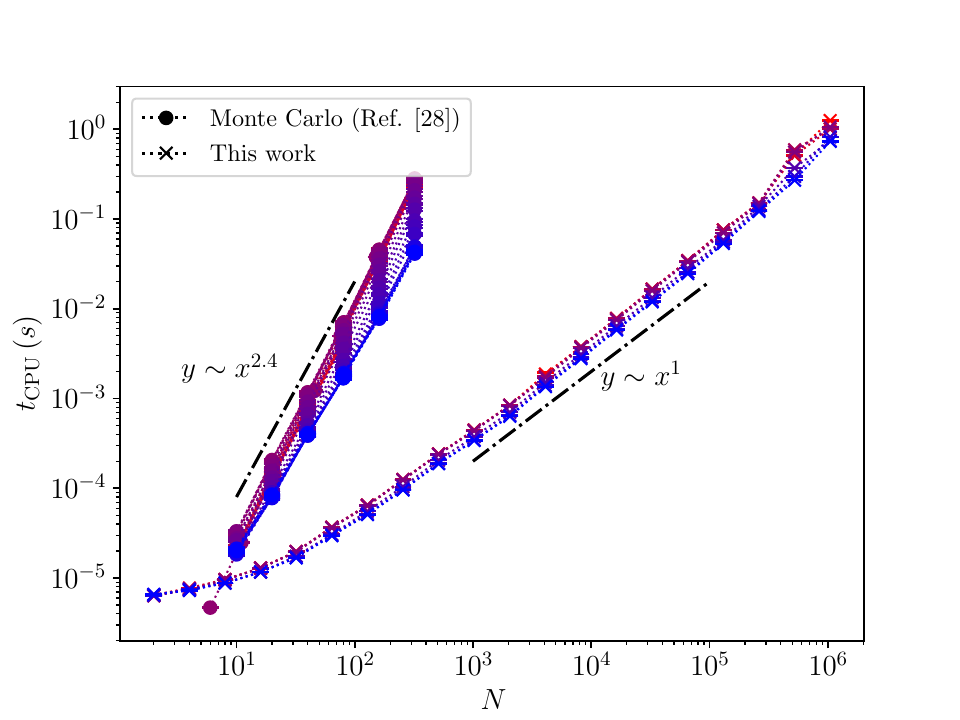}
\caption{
Mean CPU-time (in seconds, s) to obtain a random tree according to Pr\"ufer sampling ($\bullet$) {\it vs.} equilibration time by using the Monte Carlo scheme of Ref.~\cite{Amoebapaper2024} ($\times$).
Lines serve as a guide for the eye.
Different colors correspond to different values of the branching chemical potential $\mu$ (colorcode is as in Fig.~\ref{fig:n3+Rg2_usVSDJ}).
The power-law close to MC times has been extensively described in the same reference~\cite{Amoebapaper2024}.
}
\label{fig:PrueferVSAmoeba}
\end{figure}
\begin{figure}
\includegraphics[width=0.50\textwidth]{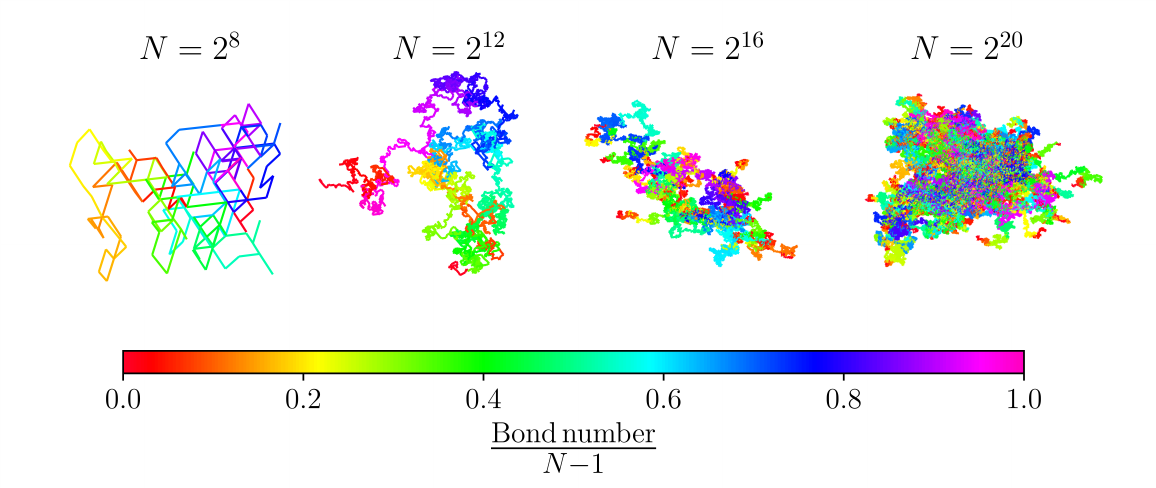}
\caption{
Trees of different sizes $N$ with Pr\"ufer-sampled connectivities which are randomly embedded on the $3d$ FCC lattice.
Bonds are shown with a colorcode indicating when they are defined within the Pr\"ufer code.
}
\label{fig:ModelConformations}
\end{figure}
\begin{figure}
\includegraphics[width=0.50\textwidth]{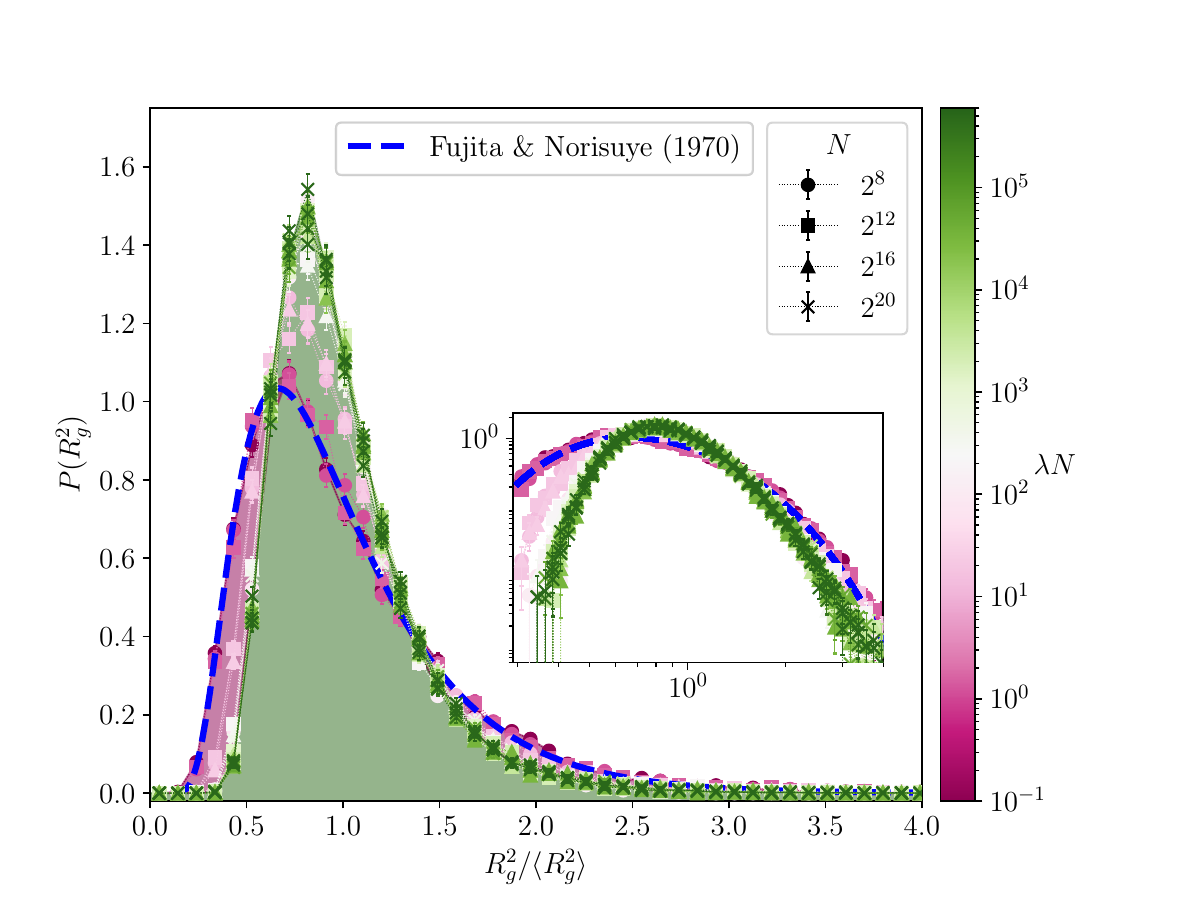}
\caption{
Probability distribution function, $P(R_g^2)$, of tree square gyration radius, $R_g^2$.
Different symbols are for different $N$ (see legend), different colors are for different values of effective tree size $\lambda N$ (colorcode on the right) which depends on $\mu$ through Eq.~\eqref{eq:LambdaFromMu}.
In the linear chain limit $\lambda N \ll1$ data agree with the exact expression (dashed line) by Fujita and Norisuye~\cite{FujitaNorisuye1970}.
(Inset)
Same data in log-log representation.
}
\label{fig:RgPDF}
\end{figure}
%

{\it Pr\"ufer sampling for random trees} -- 
In spite of the availability of these analytic results, it remains interesting to sample conformations and configurations of ideal trees.
Not all tree observables can be directly expressed as a function of the branching entropy, neither is it straightforward to express distribution functions for those whose mean values are accessible. 
Unfortunately, typical MC simulations require ${\mathcal O}(N^{2.4})$ MC steps to generate independent tree configurations, even when one employs a generalized scheme with a maximally efficient cutting and pasting of leaves~\cite{Amoebapaper2024}.
Similarly, apparently better strategies~\cite{MadrasJPhysA1992} featuring the displacement of whole branches are likely to suffer from the same difficulties as the simpler leaf-mover algorithm when introducing controlled branching activity in the polymer model (see Discussion in~\cite{Amoebapaper2024}).

The beauty of Pr\"ufer's scheme is that it allows to generate random trees in ${\cal O}(N^1)$ operations, {\it i.e.} with a computational effort proportional to the one required for the trivial sampling of linear random walks (Fig.~\ref{fig:PrueferVSAmoeba}). 
All that is required is the decoding (Sec.~S1B 
in SM~\cite{SMnote}) of a random Pr\"ufer code~\cite{Wang2009} for a random tree composition, $\{ N_f \} $, drawn from the distribution $p_{N, \mu}(N_3)$ defined in Eq.~\eqref{eq:p(n_3)}.
The tree is then embedded on the $3d$ face-centered-cubic (FCC) lattice as the following: we give one node (say $i=1$) a location on the lattice and then assign random orientations to all bonds by following the sampled tree connectivity. 
Pr\"ufer sampling allowed us (Fig.~\ref{fig:PrueferVSAmoeba}) to generate ideal trees of $N = 2^{20} \simeq 10^6$ nodes in about $1$ second of CPU-time compared to an estimated CPU-time of $\simeq 10^8 \, {\rm s} \simeq 3$ years for MC (see Fig.~\ref{fig:ModelConformations} for examples of tree conformations for various $N$).
As an application, we have used Pr\"ufer sampling to get accurate predictions for distribution functions, $P(R_g^2)$, of the tree square gyration radius.
Plots of $P(R_g^2)$ as a function of the rescaled variable $R_g^2 / \langle R_g^2\rangle$ and for different values of effective tree size $\lambda N$ (Fig.~\ref{fig:RgPDF}) span the full crossover from the linear regime (for which a computable expression exists~\cite{FujitaNorisuye1970}, dashed line) to the branched regime, where data from different systems fall nicely onto a single, universal master curve.
As a second application, we computed the variance of the square-gyration radius, ${\rm var}(R_g^2)$ (see Table~S1 
in SM~\cite{SMnote}).
In particular, its normalized value to $\langle R_g^2\rangle^2$ in the large-tree limit ($\approx 0.13$) defines the variation in the elastic free energy of the polymer due to a variation of the radius of gyration~\cite{DaoudJoanny1981}, {\it i.e.} $\delta F / (k_BT) \approx \delta R^2 / (0.13\langle R_g^2\rangle)$.

\begin{figure}
\includegraphics[width=0.50\textwidth]{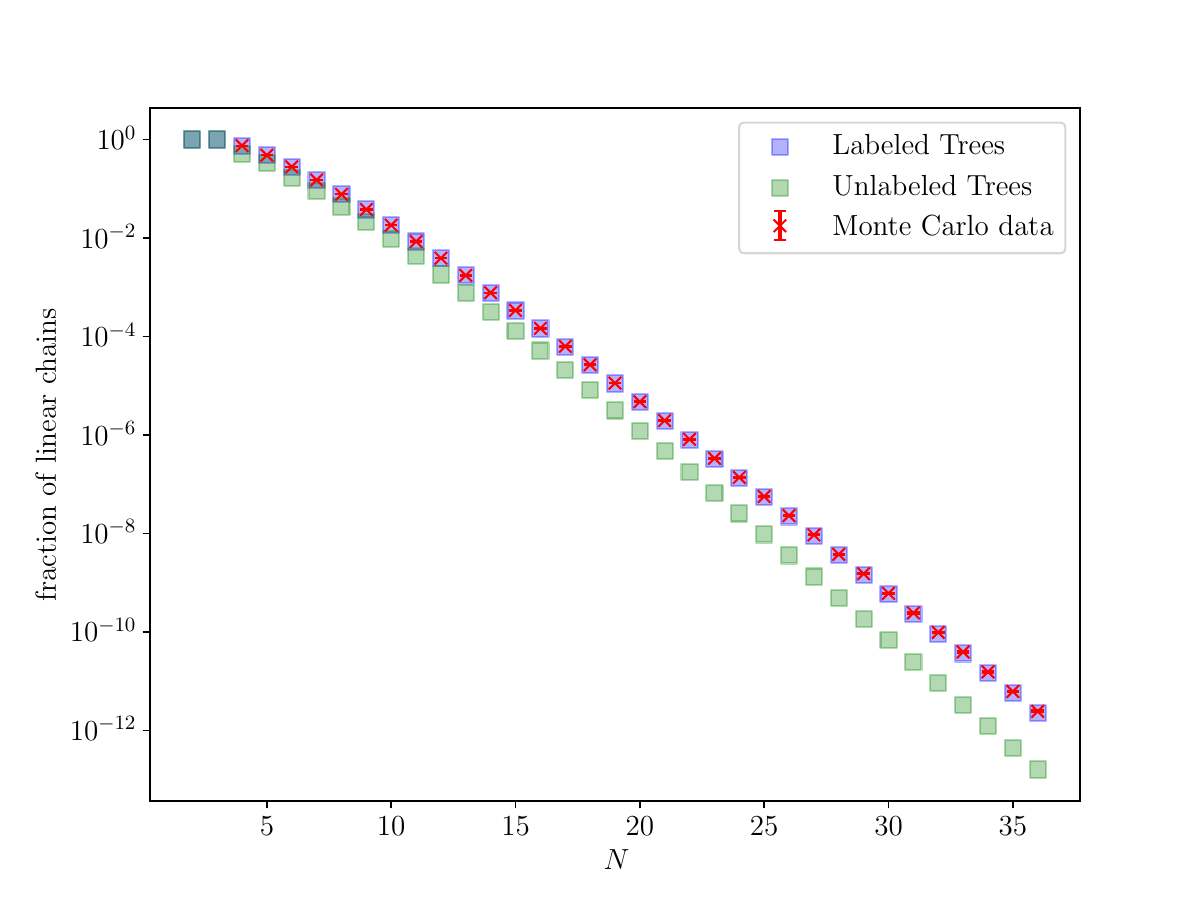}
\caption{
Fraction of linear chains with respect to all trees in the ensemble.
Results for:
labelled trees (exact, blue symbols),
unlabelled trees (exact, green symbols)
and
randomly branching polymers on the $3d$ FCC lattice (MC computer simulations, red symbols).
}
\label{fig:Labeled_vs_Unlabeled}
\end{figure}
%

{\it Labelled vs. unlabelled trees} -- 
So far we have implicitly assumed that randomly branching polymers follow the statistics of {\it labelled} trees, an assumption which is strongly supported by the striking agreement (Fig.~\ref{fig:Entropies}) between the entropies of Pr\"ufer- and MC-generated trees. 
But can we also show the reverse, namely that our generic randomly branching polymer model does {\it not} follow the statistics of {\it unlabelled} trees?
And how big is the difference between the two in the first place?

Contrary to labelled trees, the exact total number $\Omega_N^{({\rm unlabelled})}$ of {\it unlabelled} trees or distinct graph topologies of $N$ nodes is only known to some finite $N$ (for the complete list up to $N=1000$, see~\cite{UnlabelledTreesWeb}) with~\cite{Otter1948} 
\begin{equation}\label{eq:Otter}
\Omega_N^{({\rm unlabelled})} \simeq 0.5349485 \, \frac{(2.9557649)^N}{N^{5/2}}
\end{equation}
believed to be an excellent numerical approximation. 
In general, counting and sampling unlabelled trees requires more sophisticated algorithms~\cite{Nijenhuis1978} than those presented here.
On first sight, the difference in the number of labelled and unlabelled trees is enormous.
For instance, for $N=20$ the ratio $\Omega_N^{({\rm labelled})} / \Omega_N^{({\rm unlabelled})} \simeq 3\cdot 10^{17}$.
However, much of this difference is due to the experimentally irrelevant labeling entropy and has nothing to do with branching~\cite{Flory2}.
Normalizing Cayley's expression Eq.~\eqref{eq:Cayley} for the total number of labelled trees by the number of possibilities Eq.~\eqref{eq:OmegaLinearChains} to label a linear chain plus Stirling's approximation yields with
\begin{equation}\label{eq:CayleyOverLinears}
\frac{N^{N-2}}{N!/2} \simeq \sqrt{\frac2{\pi}} \, \frac{e^N}{N^{5/2}} \simeq 0.7978846 \, \frac{(2.7182818)^N}{N^{5/2}}
\end{equation}
a result which is qualitatively quite close to Eq.~\eqref{eq:Otter}.
In particular, the inverses of Eqs.~\eqref{eq:Otter} and~\eqref{eq:CayleyOverLinears} make two distinct quantitative predictions for the numerically measurable fraction of linear chains in a tree ensemble.   

To assess whether randomly branching polymers on the $3d$ FCC lattice follow the statistics of labelled or unlabelled trees, we performed MC computer simulations for trees of {\it unrestricted} functionality at varying values for a chemical potential ($\mu_1$) that controls the number of endpoints.
We then reconstructed the total number of trees up to $N=36$ by adapting the thermodynamic integration scheme already used in this work to find the fraction of linear chains with respect to all trees in the ensemble. 
Results for
unlabelled trees (using exact data from~\cite{UnlabelledTreesWeb}),
labelled trees (using the exact Cayley's formula~\eqref{eq:Cayley}),
and
MC-generated trees
are shown in Fig.~\ref{fig:Labeled_vs_Unlabeled} (green, blue and red symbols, respectively).
The perfect agreement between the exact formula for labelled trees and MC-generated data demonstrates that MC trees follow indeed labelled statistics.

{\it Discussion and conclusions} -- 
In this work we have extended results from graph theory to derive exact expressions for the composition-dependent configurational entropy of random trees and to devise a scheme for sampling trees of arbitrary branching activity in linear computing time 
by using Pr\"ufer codes.
Furthermore we have provided a first connection to the statistical theory of randomly branching polymers.

With respect to possible future perspectives, we plan to extend our approach to trees with nodes of arbitrary functionality and to introduce more chemical details like distinct identities for branching and non-branching monomers~\cite{ZimmStockmayer49} or the irreversible growth of branched polymers from a seed~\cite{Flory2,RubinsteinColby} where the statistical weight of tree connectivities depends on the number of growth paths leading to their formation.
As a first step, however, we will formally establish~\cite{DoubleFoldingPaper} the relation~\cite{RosaEveraers2019,GrosbergSoftMatter2014} between random trees and double-folded ring polymers~\cite{Ghobadpour2021,Ghobadpour2025} for which an analogue of Eq.~\eqref{eq:LambdaFromMu} was first reported~\cite{EGPhDThesis}.
Finally, Pr\"ufer codes are not the only way to encode labeled trees. 
Other ``$N-2$''-integer long codes in one-to-one correspondence with labelled trees, like for instance the so called Blob code or Dandelion code~\cite{Thompson2007}, have been proposed with possible superior features compared to our implementation of Pr\"ufer codes.
It remains therefore an interesting question to see if these codes will indeed outperform Pr\"ufer codes in the specific context of our sampling algorithm~\footnote{
Decoding bare so called Blob or Dandelion codes is also done in linear computing time, but twice as fast compared to decoding bare Pr{\"u}fer codes~\cite{Thompson2007}. This is in part because a tree can only be reconstructed from a Pr{\"u}fer code, once the functionality $f_i$ of each node $i$ is known. When the Pr{\"u}fer code is the only information available, those functionalities can only be obtained by first reading the entire Pr{\"u}fer code itself, which is an $O(n)$-operation. For Blob or Dandelion codes this step is not necessary, because information about the tree-connectivity in these codes is stored {\it locally}. A first-step attempt in improving our sampling algorithm even further might thus be by adopting Blob or Dandelion codes rather than Pr{\"u}fer codes. However, it must be emphasized that for any sampling algorithm analogous to the one presented here, the functionalities of individual nodes need to be sampled first before a code (Dandelion, Blob or Pr{\"u}fer) can be sampled. In our setup, we therefore save a step $O(n)$ of computational complexity in decoding Pr{\"u}fer codes compared to the notion used in~\cite{Thompson2007}.
}.

{\it Acknowledgements} --
We acknowledge E. Ghobadpour, M. Kolb and D. Marcato for useful discussions.
PHWvdH acknowledges financial support from PNRR\_M4C2I4.1.\_DM351 funded by NextGenerationEU and the kind hospitality of the ENS-Lyon.
AR acknowledges financial support from PNRR Grant CN\_00000013\_CN-HPC, M4C2I1.4, spoke 7, funded by Next Generation EU.

\bibliography{biblio}

\widetext
\clearpage
\begin{center}
\textbf{\Large Supplemental Material \\ \vspace*{1.5mm} The configurational entropy of random trees} \\
\vspace*{5mm}
Pieter H. W. van der Hoek, Angelo Rosa, Ralf Everaers
\vspace*{10mm}
\end{center}

\setcounter{equation}{0}
\setcounter{figure}{0}
\setcounter{table}{0}
\setcounter{page}{1}
\setcounter{section}{0}
\setcounter{page}{1}
\makeatletter
\renewcommand{\theequation}{S\arabic{equation}}
\renewcommand{\thefigure}{S\arabic{figure}}
\renewcommand{\thetable}{S\arabic{table}}
\renewcommand{\thesection}{S\arabic{section}}
\renewcommand{\thepage}{S\arabic{page}}

\tableofcontents

\clearpage

\section{On the connection between random trees and Pr\"ufer codes}\label{sec:Trees+PrueferCodes}

\subsection{From a tree to a Pr\"ufer code}\label{sec:RBP->Pruefer}
\begin{center}

\begin{tabular}{| c | c | m{7.0cm} |}

\hline

\textbf{Connectivity graph} & \textbf{Pr\"ufer code} & \textbf{Step} \\

\hhline{|=|=|=|}

\adjustbox{valign=c}{ 
    \begin{tikzpicture}
        \node[circle,draw,fill=green!20,inner sep=2pt] (5) at (0.0,0.0) {5};
        \node[circle,draw,fill=orange!20,inner sep=2pt] (4) at  (-0.75,0.43){4};
        \node[circle,draw,fill=orange!20,inner sep=2pt] (2) at (-1.5,0){2};
        \node[circle,draw,fill=green!20,inner sep=2pt] (1) at (-2.25,0.43)  {1};
        \node[circle,draw,fill=orange!20,inner sep=2pt] (6) at (-1.5,-0.75){6};
        \node[circle,draw,fill=green!20,inner sep=2pt] (3) at  (-1.5,-1.5){3};
        
        	\node[below right of = 3] (G) {}; 
	\node[above right of = 1] (H) {};

        \draw[thick,blue] (2) -- (1);
        \draw[thick,blue] (6) -- (3);
        \draw[thick,blue] (5) -- (4);
        \draw[thick,blue] (4) -- (2);
         \draw[thick,blue] (2) -- (6);

        \draw[thick, dashed,violet] (1) circle(0.4); 
     
    \end{tikzpicture}}
     & \{\} &\parbox[c][][c]{7.0cm} {\small
     For a given connectivity graph, start by separating nodes of functionality $f=1$ (the ``leaves'', the green circles) from those with $f>1$ (the orange circles).
     Then, identify the leaf $j$ with the lowest label (``$j=1$'' in this particular case, encircled). 
     } \\
     
\hline

    \adjustbox{valign=c}{
    \begin{tikzpicture}
        \node[circle,draw,fill=green!20,inner sep=2pt] (5) at (0.0,0.0) {5};
        \node[circle,draw,fill=orange!20,inner sep=2pt] (4) at  (-0.75,0.43){4};
        \node[circle,draw,fill=orange!20,inner sep=2pt] (2) at (-1.5,0){2};
        \node[circle,draw,fill=blue!20,inner sep=2pt] (1) at (-2.25,0.43)  {1};
        \node[circle,draw,fill=orange!20,inner sep=2pt] (6) at (-1.5,-0.75){6};
        \node[circle,draw,fill=green!20,inner sep=2pt] (3) at  (-1.5,-1.5){3};
        
        	\node[below right of = 3] (G) {};
	\node[above right of = 1] (H) {};

        \draw[thick,red] (2) -- (1);
        \draw[thick,blue] (6) -- (3);
        \draw[thick,blue] (5) -- (4);
        \draw[thick,blue] (4) -- (2);
         \draw[thick,blue] (2) -- (6);

        \draw[thick, dashed,violet] (3) circle(0.4); 

    \end{tikzpicture}}
    & \{\textcolor{red}{2}\}&\parbox[c][][c]{7.0cm} {\small
    The first entry in the Pr\"ufer code is the label $i$ of the node to which $j$ is attached (``$j=2$'' in this particular case).
    To proceed, remove $j$ from the list of leaves, then reduce the functionality $f_i$ of node $i$ by one, and update the identity $j$ of the leaf having the lowest label among the present leaves (``$j=3$'' in this particular case, encircled).
} \\

\hline

\adjustbox{valign=c}{
    \begin{tikzpicture}
        \node[circle,draw,fill=green!20,inner sep=2pt] (5) at (0.0,0.0) {5};
        \node[circle,draw,fill=orange!20,inner sep=2pt] (4) at  (-0.75,0.43){4};
        \node[circle,draw,fill=orange!20,inner sep=2pt] (2) at (-1.5,0){2};
        \node[circle,draw,fill=blue!20,inner sep=2pt] (1) at (-2.25,0.43)  {1};
        \node[circle,draw,fill=green!20,inner sep=2pt] (6) at (-1.5,-0.75){6};
        \node[circle,draw,fill=blue!20,inner sep=2pt] (3) at  (-1.5,-1.5){3};
        
        	\node[below right of = 3] (G) {};
	\node[above right of = 1] (H) {};

        \draw[thick] (2) -- (1);
        \draw[thick,red] (6) -- (3);
        \draw[thick,blue] (5) -- (4);
        \draw[thick,blue] (4) -- (2);
         \draw[thick,blue] (2) -- (6);

        \draw[thick, dashed,violet] (5) circle(0.4); 
                                      
    \end{tikzpicture}}
    &\{2,\textcolor{red}{6}\} &\parbox[c][][c]{7.0cm} {\small
    As in the first step, the second entry in the Pr\"ufer code is the label $i$ of the node to which the updated $j$ is attached (``$i=6$'' in this particular case).
    Otherwise proceed as before, but note that now the functionality $f_i$ (``$f_6$'' in this particular case) of node $i$ is reduced to one so that it needs to be counted among the leaves.
    } \\
    
\hline

\adjustbox{valign=c}{
    \begin{tikzpicture}
        \node[circle,draw,fill=blue!20,inner sep=2pt] (5) at (0.0,0.0) {5};
        \node[circle,draw,fill=green!20,inner sep=2pt] (4) at  (-0.75,0.43){4};
        \node[circle,draw,fill=orange!20,inner sep=2pt] (2) at (-1.5,0){2};
        \node[circle,draw,fill=blue!20,inner sep=2pt] (1) at (-2.25,0.43)  {1};
        \node[circle,draw,fill=green!20,inner sep=2pt] (6) at (-1.5,-0.75){6};
        \node[circle,draw,fill=blue!20,inner sep=2pt] (3) at  (-1.5,-1.5){3};
        
        	\node[below right of = 3] (G) {};
	\node[above right of = 1] (H) {};

        \draw[thick] (2) -- (1);
        \draw[thick] (6) -- (3);
        \draw[thick,red] (5) -- (4);
        \draw[thick,blue] (4) -- (2);
         \draw[thick,blue] (2) -- (6);

        \draw[thick, dashed,violet] (4) circle(0.4); 

    \end{tikzpicture}}
&  \{2,6,\textcolor{red}{4}\} & \small{
Continue as before.
} \\

\hline

\adjustbox{valign=c}{
    \begin{tikzpicture}
        \node[circle,draw,fill=blue!20,inner sep=2pt] (5) at (0.0,0.0) {5};
        \node[circle,draw,fill=blue!20,inner sep=2pt] (4) at  (-0.75,0.43){4};
        \node[circle,draw,fill=green!20,inner sep=2pt] (2) at (-1.5,0){2};
        \node[circle,draw,fill=blue!20,inner sep=2pt] (1) at (-2.25,0.43)  {1};
        \node[circle,draw,fill=green!20,inner sep=2pt] (6) at (-1.5,-0.75){6};
        \node[circle,draw,fill=blue!20,inner sep=2pt] (3) at  (-1.5,-1.5){3};
        
        	\node[below right of = 3] (G) {};
	\node[above right of = 1] (H) {};

        \draw[thick] (2) -- (1);
        \draw[thick] (6) -- (3);
        \draw[thick] (5) -- (4);
        \draw[thick,red] (4) -- (2);
         \draw[thick,blue] (2) -- (6);

    \end{tikzpicture}}
    & \{2,6,4,\textcolor{red}{2}\} &\parbox[c][][c]{7.0cm} {\small
    There is no choice to be coded for the last bond as there are only two leaves left (labels ``$2$'' and ``$6$'' in this particular case) when the Pr\"ufer code has reached a length of $N-2$ entries.
    Note that at the end of the procedure each node appears ``$f-1$'' times in the completed code.
    } \\
    
\hline

\end{tabular}

\end{center}

\newpage

\subsection{From a Pr\"ufer code to a tree}\label{sec:Pruefer->RBP}

\begin{center}

\begin{tabular}{| c | c | m{7.0cm} |}

\hline

\textbf{Connectivity graph} & \textbf{Pr\"ufer code} & \textbf{Step}\\

\hhline{|=|=|=|}

\adjustbox{valign=c}{%
	\begin{tikzpicture}
        \node[circle,draw,fill=green!20,inner sep=2pt] (5) at (1.5,0) {5};
        \node[circle,draw,fill=orange!20,inner sep=2pt] (4) at (0.5,0) {4};
        \node[circle,draw,fill=orange!20,inner sep=2pt] (2) at (-1.5,0) {2};
        \node[circle,draw,fill=green!20,inner sep=2pt] (1) at (-2.5,0) {1};
        \node[circle,draw,fill=orange!20,inner sep=2pt] (6) at (2.5,0) {6};
        \node[circle,draw,fill=green!20,inner sep=2pt] (3) at (-0.5,0) {3};
        
	\node[below right of = 3] (G) {};
	\node[above right of = 1] (H) {};
	\node[below right of = G] (I) {};
	\node[above right of = H] (J) {};

        \draw[thick, dashed] (1) -- ++(-0,-0.75);  
        \draw[thick, dashed] (2) -- ++(-0.75,0.43); 
        \draw[thick, dashed] (2) -- ++(0.75,0.43);  
        \draw[thick, dashed] (2) -- ++(-0,-0.75);
        \draw[thick, dashed] (3) -- ++(0,-0.75);   
        \draw[thick, dashed] (4) -- ++(0,-0.75);   
        \draw[thick, dashed] (4) -- ++(0,0.75);   
        \draw[thick, dashed] (5) --++(-0,-0.75);  
        \draw[thick, dashed] (6)-- ++(0,0.75);     
        \draw[thick, dashed] (6)-- ++(0,-0.75);     
        
        \draw[thick, dashed,violet] (1) circle(0.4);

    \end{tikzpicture}}
   & \{2,6,4,2\} & \parbox[c][][c]{7.0cm} {\small
   Start by ordering the tree nodes, based on their functionality. The functionality of a node can be read off from how often its label appears in the Pr\"ufer code.
   Each node $i$ appears $f_i -1$ times, {\it i.e.} nodes whose indices do not appear are ``$f=1$''-functional nodes (or ``leaves'', in green in the figure).
   Identify the leaf with the lowest label $j$ (``$j=1$'' in this particular case, encircled).
   }\\ 
   
\hline

    \adjustbox{valign=c}{
    \begin{tikzpicture}
        \node[circle,draw,fill=green!20,inner sep=2pt] (5) at (1.5,0.0) {5};
        \node[circle,draw,fill=orange!20,inner sep=2pt] (4) at (0.5,0) {4};
        \node[circle,draw,fill=orange!20,inner sep=2pt] (2) at (-1.75,0) {2};
        \node[circle,draw,fill=blue!20,inner sep=2pt] (1) at (-2.5,0.43) {1};
        \node[circle,draw,fill=orange!20,inner sep=2pt] (6) at (2.5,0) {6};
        \node[circle,draw,fill=green!20,inner sep=2pt] (3) at (-0.5,0) {3};
        	\node[] (G) at (-2.5,-0.43){};
	\node[above right of = 1] (H) {};
	\node[below right of = G] (Z) {};

        \draw[thick, dashed] (2) -- ++(0.75,0.43);  
         \draw[thick, dashed] (2) -- ++(-0,-0.75);
        \draw[thick, dashed] (3) -- ++(0,-0.75);   
        \draw[thick, dashed] (4) -- ++(0,-0.75);   
        \draw[thick, dashed] (4) -- ++(0,0.75);   
        \draw[thick, dashed] (5) --++(-0,-0.75);  
        \draw[thick, dashed] (6)-- ++(0,0.75);     
        \draw[thick, dashed] (6)-- ++(0,-0.75);     
        
        \draw[thick,red] (2) -- (1);
        
        \draw[thick, dashed,violet] (3) circle(0.4); 
    \end{tikzpicture}}
    &  \{\textcolor{red}{2},6,4,2\} & \parbox[c][][c]{7.0cm} {\small
    Attach $j$ to the node $i$ identified by the first index of the Pr\"ufer code (``$i=2$'' in this particular case).
    Update the functionality of the node $i$ and identify the new leaf $j$ with the lowest index (``$j=3$'' in this particular case, encircled). 
    } \\

\hline

    \adjustbox{valign=c}{
    \begin{tikzpicture}
        \node[circle,draw,fill=green!20,inner sep=2pt] (5) at (1.5,0.0) {5};
        \node[circle,draw,fill=orange!20,inner sep=2pt] (4) at (0.5,0) {4};
        \node[circle,draw,fill=orange!20,inner sep=2pt] (2) at (-1.75,0) {2};
        \node[circle,draw,fill=blue!20,inner sep=2pt] (1) at (-2.5,0.43) {1};
        \node[circle,draw,fill=green!20,inner sep=2pt] (6) at (-0.5,-0.32) {6};
        \node[circle,draw,fill=blue!20,inner sep=2pt] (3) at (-0.5,0.43) {3};
         \node[] (G) at (-2.5,-0.43){};
	\node[above right of = 1] (H) {};
	\node[below right of = G] (Z) {};

        \draw[thick, dashed] (2) -- ++(0.75,0.43);  
         \draw[thick, dashed] (2) -- ++(-0,-0.75);
        \draw[thick, dashed] (4) -- ++(0,-0.75);   
        \draw[thick, dashed] (4) -- ++(0,0.75);   
        \draw[thick, dashed] (5) --++(-0,-0.75);  
        \draw[thick, dashed] (6)-- ++(0,-0.75);     
                
        \draw[thick] (2) -- (1);
        \draw[thick,red] (6) -- (3);
        
       \draw[thick, dashed,violet] (5) circle(0.4); 

    \end{tikzpicture}}
    &  \{2,\textcolor{red}{6},4,2\} &\parbox[c][][c]{7.0cm} {\small
    Attach $j$ to the node $i$ identified by the next index of the Pr\"ufer code (``$i=6$'' in this particular case).
    Proceed as before and note that when a ``$f=2$''-node is connected to a leaf, it becomes itself a leaf. 
    } \\

\hline

    \adjustbox{valign=c}
    {\begin{tikzpicture}
        \node[circle,draw,fill=blue!20,inner sep=2pt] (5) at (0.5,0.43) {5};
        \node[circle,draw,fill=green!20,inner sep=2pt] (4) at (0.5,-0.32) {4};
        \node[circle,draw,fill=orange!20,inner sep=2pt] (2) at (-1.75,0) {2};
        \node[circle,draw,fill=blue!20,inner sep=2pt] (1) at (-2.5,0.43) {1};
        \node[circle,draw,fill=green!20,inner sep=2pt] (6) at (-0.5,-0.32) {6};
        \node[circle,draw,fill=blue!20,inner sep=2pt] (3) at (-0.5,0.43) {3};
        
        \node[below right of = 2] (G) {};
	\node[above right of = 1] (H) {};
	\node[below left of = G] (I) {};

        \draw[thick, dashed] (2) -- ++(0.75,0.43);  
         \draw[thick, dashed] (2) -- ++(-0,-0.75);
        \draw[thick, dashed] (4) -- ++(0,-0.75);  
        \draw[thick, dashed] (6)-- ++(0,-0.75);     
        
        \draw[thick] (2) -- (1);
        \draw[thick] (6) -- (3);
        \draw[thick,red] (5) -- (4);
        
	\draw[thick, dashed,violet] (4) circle(0.4); 
    \end{tikzpicture}}
               &  \{2,6,\textcolor{red}{4},2\} & \parbox[c][][c]{7.0cm} {\small
               Continue as before.
               } \\
               
\hline

      \adjustbox{valign=c}{
      \begin{tikzpicture}
        \node[circle,draw,fill=blue!20,inner sep=2pt] (5) at (0.0,0.0) {5};
        \node[circle,draw,fill=blue!20,inner sep=2pt] (4) at (-0.75,0.43) {4};
        \node[circle,draw,fill=green!20,inner sep=2pt] (2) at (-1.5,0) {2};
        \node[circle,draw,fill=blue!20,inner sep=2pt] (1) at (-2.25,0.43) {1};
        \node[circle,draw,fill=green!20,inner sep=2pt] (6) at (1.0,-0.32) {6};
        \node[circle,draw,fill=blue!20,inner sep=2pt] (3) at (1.0,0.43) {3};
        
        	\node[below left of = 5] (G) {};
	\node[above right of = 1] (H) {};
	\node[below left of = G] (I) {};

        \draw[thick, dashed] (2) -- ++(-0,-0.75);
       	\draw[thick, dashed] (6)-- ++(0,-0.75);     
	
        \draw[thick] (2) -- (1);
        \draw[thick] (6) -- (3);
        \draw[thick] (5) -- (4);
        \draw[thick,red] (4) -- (2);

    \end{tikzpicture}}
    &  \{2,6,4,\textcolor{red}{2}\} &\parbox[c][][c]{7.0cm} {\small
    Continue as before until all ``$N-2$'' indices of the Pr\"ufer code have been read and only two leaves remain (labels ``$2$'' and ``$6$'' in this particular case). 
    } \\

\hline

       \adjustbox{valign=c}{
       \begin{tikzpicture}
        \node[circle,draw,fill=blue!20,inner sep=2pt] (5) at (0.0,0.0) {5};
        \node[circle,draw,fill=blue!20,inner sep=2pt] (4) at (-0.75,0.43) {4};
        \node[circle,draw,fill=blue!20,inner sep=2pt] (2) at (-1.5,0) {2};
        \node[circle,draw,fill=blue!20,inner sep=2pt] (1) at (-2.25,0.43) {1};
        \node[circle,draw,fill=blue!20,inner sep=2pt] (6) at (-1.5,-0.75) {6};
        \node[circle,draw,fill=blue!20,inner sep=2pt] (3) at (-1.5,-1.5) {3};
	
	\node[below right of = 3] (G) {};
	\node[above right of = 1] (H) {};
 	
        \draw[thick] (2) -- (1);
        \draw[thick] (6) -- (3);
        \draw[thick] (5) -- (4);
        \draw[thick] (4) -- (2);
        \draw[thick] (2) -- (6);

    \end{tikzpicture}}
    &  \{2,6,4,2\} &\parbox[c][][c]{7.0cm} {\small
    In the last step, connect the remaining leaves so to complete the connectivity of the tree.
    } \\
                        
\hline

\end{tabular}
               
\end{center}

\newpage

\section{Limiting behaviors of branch-nodes}\label{sec:ApproximateRBP}
In terms of the partition function $Z_{N, \mu} = \sum_{N_3=0}^{N_{3,\rm max}} \Omega_N(N_3) \, e^{\beta\mu N_3}$ (see Eqs.~(8) and~(10) 
in main text)
the mean number of branch-nodes is given by
\begin{equation} \label{eq:Exp_branchpoints}
\langle N_3 \rangle = \frac{\partial (\log(Z_{N, \mu}))}{\partial (\beta \mu)}  = \frac{ \sum_{N_3=0}^{N_{3,\rm max}} e^{-\beta F_{N, \mu}(N_3)} \, N_3 }{ \sum_{N_3=0}^{N_{3,\rm max}} e^{-\beta F_{N, \mu}(N_3)} } \, ,
\end{equation}
where
\begin{equation}\label{eq:branching_free_energy}
\beta F_{N, \mu}(N_3) = -\log(\Omega_N(N_3)) - \beta\mu N_3
\end{equation}
is the branching {\it free energy}.
We consider the following two limit behaviors:
\begin{itemize}
\item
In the limit where the polymer ensemble is constituted only by the two states with $N_3 = 0$ and $N_3 =1$, Eq.~\eqref{eq:Exp_branchpoints} is given by:
\begin{equation}\label{eq:branching_probability_N3=0andN3=1}
\langle N_3 \rangle = \frac{ 0 \cdot \Omega_N(N_3 = 0) + 1 \cdot \Omega_N(N_3 = 1) \, e^{\beta \mu}}{\Omega_N(N_3 = 0) + \Omega_N(N_3 = 1) \, e^{\beta \mu}} = \frac{ \frac{(N-2)(N-3)}{3!} \, e^{\beta\mu} }{ 1 + \frac{(N-2)(N-3)}{3!} \, e^{\beta\mu} } \, .
\end{equation}
In particular, in the limits where $e^{\beta\mu} \ll 0$ and $N\gg 1$, Eq.~\eqref{eq:branching_probability_N3=0andN3=1} reads:
\begin{equation}\label{eq:branching_probability_trees_small}
\langle N_3 \rangle \simeq \frac{N^2}6 e^{\beta \mu} = \frac13 (e^{(\beta\mu/2-\log(2))/2} \, N)^2 \, .
\end{equation}
\item
By taking $N \gg 1$ also $N_3 \gg 1$ and by using the Stirling asymptotic formula for the factorial, Eq.~\eqref{eq:branching_free_energy} can be approximated by:
%
\begin{equation}\label{eq:branching_free_energy_approx}
\beta F_{N, \mu}(N_3) \simeq - (\beta \mu + \log(2)) N_3 - N\log(N) + N 
- 2N_3\log\left( \frac{N}{2N_3} \right) - (N-2N_3)\log\left( \frac{N}{N-2N_3} \right) \, .
\end{equation}
%
By noticing that the last two terms of Eq.~\eqref{eq:branching_free_energy_approx} represent the logarithmic approximation of a binomial distribution which, by the central limit theorem, can be approximated by a normal distribution with mean value around its maximum and by solving for $\left. \frac{\partial (\beta F_{N,\mu})}{\partial N_3} \right|_{N_3=\langle N_3\rangle}= 0$, we get:
\begin{equation}\label{eq:branching_probability_trees_big}
\left. \langle N_3 \rangle \right|_{N\gg 1} = \frac{N}{2 + e^{-(\beta\mu - \log(2))/2}} \, .
\end{equation}
\end{itemize}
Otherwise, the full solution of Eq.~\eqref{eq:Exp_branchpoints} has to be carried out numerically.
Results for different $N$ and $\mu$ are shown in Fig.~2(a) 
in main text (``$\blacksquare$'' symbols) in comparison to results (``$\times$'' symbols) for Pr\"ufer-sampled trees (see also Fig.~\ref{fig:n3+Rg2-test}, top panel).
Notice, in particular, the two regimes Eqs.~\eqref{eq:branching_probability_trees_small} and~\eqref{eq:branching_probability_trees_big}.

\section{Exact mean-square gyration radius of branched polymers and the Kramers' theorem}\label{sec:Exact<Rg2>}
In this section we revisit the famous Kramers' theorem~\cite{Kramers1946}, and use it in connection with our theory to compute the mean-square gyration radius, $\langle R_g^2\rangle$, for ideal randomly branching polymers.

For a single polymer configuration,
\begin{equation}\label{eq:Gyration_Radius}
R_g^2 = \frac1{2N^2} \sum_{ij} {\vec r}_{ij}^{\, 2} \, , 
\end{equation}
where ${\vec r}_{ij} \equiv {\vec r}_i - {\vec r}_j$ is the spatial distance between nodes $i$ and $j$ in the polymer. 
In a branched polymer without cycles, any pair of nodes $i$ and $j$ is uniquely connected by a linear path of $n<N$ bonds. 
In ideal conditions these $n$ bonds behave like a random walk, {\it i.e.} $\langle {\vec r}_{ij}^{\, 2} \rangle = n$.

The Kramers' theorem~\cite{Kramers1946} provides a clever relation to obtain $\langle R_g^2\rangle$ by counting the total number of different ways the tree can be divided into two parts, namely how often individual bonds are part of a path between all pairs of nodes $i,j$. 
The reasoning proceeds as follows.
Each bond represents a potential {\it cut}, that splits the polymer in $n$ and $N-n$ nodes.
By summing over all possible choices for $i,j$, each bond contributes $2 n (N-n)$ times to the sum in Eq.~\eqref{eq:Gyration_Radius}. 
By that, the square gyration radius of an individual polymer can be expressed as
%
$ R_g^2 = \frac1{N^2} \sum_{\rm cuts} n (N-n) $
%
where the summation $\sum_{\rm{cuts}}$ is over all $N-1$ bonds of a polymer.
This expression can be then averaged on an ensemble of multiple polymer conformations, {\it i.e.}
\begin{equation}\label{eq:Gyration_Radius3}
\langle R_g^2 \rangle = \frac{N-1}{N^2} \sum_{\rm cuts} \langle n (N-n) \rangle \, ,
\end{equation}
that, in essence, constitutes the Kramers' theorem and that, equivalently, can be also written as 
\begin{equation}\label{eq:Rg_result}
\langle R_g^2 \rangle = \frac{N-1}{N^2} \, \sum_{n=1}^{N-1} n(N-n) \, p_{N,\mu}(n) \, ,
\end{equation}
where $p_{N,\mu}(n)$ represents the probability of getting a sub-tree of weight $=n<N$ in a tree of total weight $=N$.

To construct such probability, we introduce the new quantities
\begin{eqnarray}
\omega_N(N_3) & \equiv & N \, \Omega_N(N_3) \, , \label{eq:RootedLabelledTreeTotalNumber} \\
\omega_N^{(f)}(N_3) & \equiv & N_f \, \Omega_N(N_3) \, , \label{eq:RootedLabelledTreeTotalNumber-fFunct}
\end{eqnarray}
that represent, respectively, the total number of {\it rooted} labelled tree~\footnote{A rooted tree is a tree with a special node called the {\it root}, which serves as a starting point for all other nodes in the tree.} and the total number of {\it rooted} labelled tree at $f$-functional sites.
Then, by also introducing the corresponding partition function
\begin{equation}\label{eq:RootedLabelledTreeTotalNumber-PartFunct}
\zeta_{N,\mu}^{(f)} \equiv \sum_{n_3} \omega_N^{(f)}(N_3) \, e^{\beta\mu N_3} \, , 
\end{equation}
the probability $p_{N,\mu}(n)$ is finally given by the expression:
\begin{equation}\label{eq:p(n)}
p_{N,\mu}(n) = \frac{ \frac{N!}{n! \, (N-n)!} \, \left( \zeta_{n,\mu}^{(1)} + e^{\beta\mu} \, \zeta_{n,\mu}^{(2)} \right) \, \left( \zeta_{N-n,\mu}^{(1)} + e^{\beta\mu} \, \zeta_{N-n,\mu}^{(2)} \right) }{ \sum_{n=1}^{N-1} \frac{N!}{n! \, (N-n)!} \, \left( \zeta_{n,\mu}^{(1)} + e^{\beta\mu} \, \zeta_{n,\mu}^{(2)} \right) \, \left( \zeta_{N-n,\mu}^{(1)} + e^{\beta\mu} \, \zeta_{N-n,\mu}^{(2)} \right) } \, ,
\end{equation}
where the binomial factors in Eq.~\eqref{eq:p(n)} express ``the mixing entropy of labels'', namely the total number of possible ways of dividing $N$ labels over two sub-trees of $n$ and $N-n$ nodes.

Numerical evaluation of $\langle R_g^2\rangle$ by using Eqs.~\eqref{eq:Rg_result}-\eqref{eq:p(n)} and for different $N$ and $\mu$ are shown in Fig.~2(b) 
in main text (``$\blacksquare$'' symbols) in comparison to results (``$\times$'' symbols) for Pr\"ufer-sampled trees (see also Fig.~\ref{fig:n3+Rg2-test}, bottom panel).
Notice, in particular, the expected~\cite{RubinsteinColby} behaviors $\langle R_g^2 \rangle \sim N$ and $\langle R_g^2 \rangle \sim N^{1/2}$ valid, respectively, in the low-branching and high-branching regime and for $N\gg 1$.

\section{de Gennes-Daoud-Joanny theory of randomly branching polymers}\label{sec:ReviewingDJ}
de Gennes first in 1968~\cite{DeGennes1968} and later on in 1981 Daoud and Joanny~\cite{DaoudJoanny1981} derived the following field-theoretic expression for the partition function of branched polymers:
\begin{equation}\label{eq:DJ3}
Z_{N, \lambda}^{\rm deGDJ} = \frac{I_1 \! \left( 2 \lambda N \right)}{\lambda N} \simeq \left\{ \begin{array}{cc} 1 + \frac{(\lambda N)^2}2 \, , & \lambda N \ll 1 \\ \\ \frac{e^{2\lambda N}}{2\sqrt{\pi}(\lambda N)^{3/2}} \, , & \lambda N \gg 1 \end{array} \right. \, , 
\end{equation}
where $I_1(x)$ is the modified Bessel function of the first kind and $\lambda$ (or rather, $\lambda^2$) is a phenomenological ``control'' parameter that defines the ``activity'' of the ``$f=3$"-functional nodes.
In terms of the partition function Eq.~\eqref{eq:DJ3}, Daoud and Joanny gave computable expressions for $\langle N_3\rangle$ and $\langle R_g^2\rangle$ that we summarize below.

By using the formula (see Eq.~(A.7) in~\cite{DaoudJoanny1981}) for the mean number of branch-nodes,
\begin{equation}\label{eq:<n3>DJ}
\langle N_3 \rangle =  \lambda N \, \frac{\partial \log(Z_{N, \lambda}^{\rm deGDJ})}{\partial(2 \lambda N)} \, ,
\end{equation}
we get:
\begin{equation}\label{eq:DJ1}
\langle N_3\rangle \simeq
\left\{
\begin{array}{lc}
\frac12 (\lambda N)^2 \, , & \lambda N \ll 1 \\
\\
\lambda N \, , & \lambda N \gg 1
\end{array}
\right. .
\end{equation}
%
For $\lambda N \gg 1$, Eq.~\eqref{eq:DJ1} is equivalent to Eq.~\eqref{eq:branching_probability_trees_big} by choosing
\begin{equation}\label{eq:FudgeLambda}
\lambda^{-1} = 2 + e^{-(\beta\mu - \log(2))/2} \, .
\end{equation}
However, for the same choice of $\lambda$ and in the opposite limit $\lambda N \ll 1$, Eq.~\eqref{eq:DJ1} does not coincide with Eq.~\eqref{eq:branching_probability_trees_small}.
To reconcile this apparent discrepancy, it is useful to notice that the deGennes-Daoud-Joanny theory adopts the Gaussian, or continuum limit, approximation~\cite{DoiEdwards,RubinsteinColby} for a polymer chain.
We may easily recover this limit by replacing every bond in our model by a sequence of, say, $n$ shorter bonds and by imposing then the conservation of the mean number of branch-points:
\begin{equation}\label{eq:RenormalizeMu}
\frac{N}{2 + e^{-(\beta\mu - \log(2))/2}} \equiv \lambda N = \langle N_3 \rangle = \lambda^\ast nN \equiv \frac{nN}{e^{-(\beta\mu^\ast - \log(2))/2}}  \, ,
\end{equation}
that defines the new, renormalized branching parameter $\lambda^\ast$ in terms of an appropriate choice, $\mu=\mu^\ast$, of the microscopic parameter.

As for the mean-square gyration radius, Daoud and Joanny proposed (see Eq.~(A.12) in~\cite{DaoudJoanny1981}) the following expression adapted from the Kramers' theorem:
\begin{equation}\label{eq:<Rg2>DJ}
\langle R_g^2 \rangle = \frac1{N-1} \, \frac{ \sum_{n=1}^{N-1} n(N-n) \, Z_{n, \lambda}^{\rm deGDJ} \, Z_{N-n, \lambda}^{\rm deGDJ} }{ \sum_{n=1}^{N-1} Z_{n, \lambda}^{\rm deGDJ} \, Z_{N-n, \lambda}^{\rm deGDJ} } \, .
\end{equation}

Results for $\langle N_3\rangle$ and $\langle R_g^2 \rangle$ computed, respectively, from Eq.~\eqref{eq:<n3>DJ} and~\eqref{eq:<Rg2>DJ} and for $\lambda$ defined by Eq.~\eqref{eq:FudgeLambda} are shown as solid lines in panels (a) and (b) in Fig.~2 
in main text.


\clearpage
\section*{Supplemental figures \& table}

\clearpage
\begin{figure}
$$
\begin{array}{c}
\includegraphics[width=0.65\textwidth]{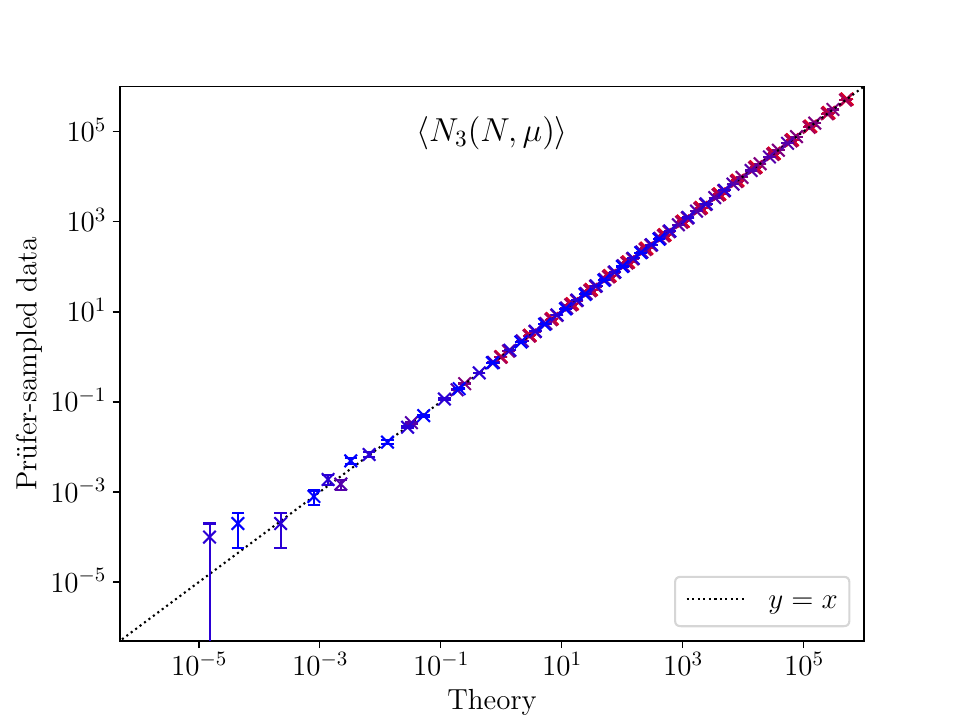} \\
\includegraphics[width=0.65\textwidth]{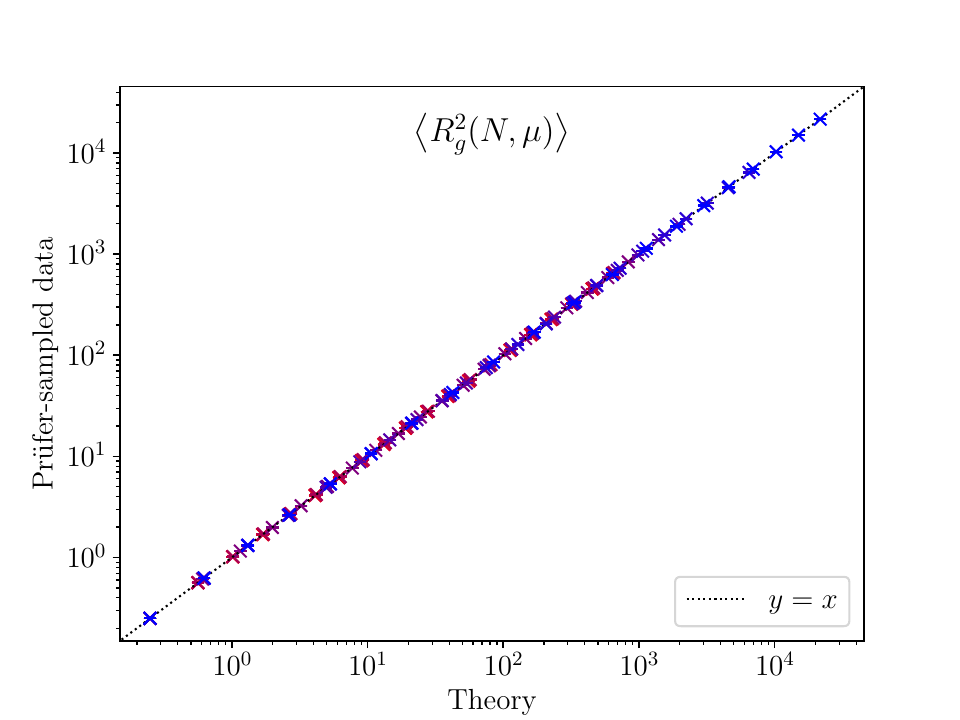}
\end{array}
$$
\caption{
Mean number of branch-nodes $\langle N_3 \rangle$ (top panel) and mean-square gyration radius $\langle R_g^2 \rangle$ (bottom panel) of lattice trees from Pr\"ufer sampling ($y$-axis) {\it vs.} analytical predictions from, respectively, Eq.~\eqref{eq:Exp_branchpoints} and Eq.~\eqref{eq:Rg_result} ($x$-axis).
We report values for different tree size $N$ and different branch potential $\beta\mu$ (colorcode is as in Fig.~2 
in main text).
The perfect alignment of Pr\"ufer sampling and analytical predictions is represented by the $y=x$ axis (dotted line).
}
\label{fig:n3+Rg2-test}
\end{figure}
%

\clearpage
\begin{table}[h!]
\centering
\begin{tabular}{cccc}
\,\,\, $N$ \,\,\, & \,\,\, $\langle R_g^2 \rangle$ \,\,\, & \,\,\, ${\rm var}(R_g^2) \equiv \langle R_g^4\rangle - \langle R_g^2 \rangle^2$ \,\,\, & \,\,\, ${\rm var}(R_g^2) / \langle R_g^2\rangle^2$ \,\,\, \\
\\
\hline
\hline
\\
$2^{ 2 }$ & 0.611 $\pm$ 0.002 & 0.039 $\pm$ 0.001 & 0.104 $\pm$ 0.001 \\
\\
$2^{ 3 }$ & 1.159 $\pm$ 0.004 & 0.177 $\pm$ 0.003 & 0.132 $\pm$ 0.002 \\
\\
$2^{ 4 }$ & 1.981 $\pm$ 0.007 & 0.56 $\pm$ 0.012 & 0.143 $\pm$ 0.003 \\
\\
$2^{ 5 }$ & 3.245 $\pm$ 0.012 & 1.587 $\pm$ 0.039 & 0.151 $\pm$ 0.003 \\
\\
$2^{ 6 }$ & 5.056 $\pm$ 0.019 & 3.67 $\pm$ 0.086 & 0.144 $\pm$ 0.003 \\
\\
$2^{ 7 }$ & 7.677 $\pm$ 0.029 & 8.347 $\pm$ 0.208 & 0.142 $\pm$ 0.003 \\
\\
$2^{ 8 }$ & 11.497 $\pm$ 0.044 & 19.269 $\pm$ 0.526 & 0.146 $\pm$ 0.003 \\
\\
$2^{ 9 }$ & 16.841 $\pm$ 0.064 & 40.977 $\pm$ 1.065 & 0.144 $\pm$ 0.003 \\
\\
$2^{ 10 }$ & 24.478 $\pm$ 0.09 & 80.297 $\pm$ 1.905 & 0.134 $\pm$ 0.003 \\
\\
$2^{ 11 }$ & 35.546 $\pm$ 0.131 & 169.92 $\pm$ 4.141 & 0.134 $\pm$ 0.003 \\
\\
$2^{ 12 }$ & 50.519 $\pm$ 0.18 & 331.602 $\pm$ 7.416 & 0.13 $\pm$ 0.002 \\
\\
$2^{ 13 }$ & 72.805 $\pm$ 0.27 & 729.226 $\pm$ 28.712 & 0.138 $\pm$ 0.005 \\
\\
$2^{ 14 }$ & 103.567 $\pm$ 0.38 & 1435.927 $\pm$ 36.667 & 0.134 $\pm$ 0.003 \\
\\
$2^{ 15 }$ & 146.592 $\pm$ 0.521 & 2801.054 $\pm$ 70.735 & 0.13 $\pm$ 0.003 \\
\\
$2^{ 16 }$ & 207.192 $\pm$ 0.743 & 5585.903 $\pm$ 155.163 & 0.13 $\pm$ 0.003 \\
\\
$2^{ 17 }$ & 295.088 $\pm$ 1.086 & 11528.461 $\pm$ 297.711 & 0.132 $\pm$ 0.003 \\
\\
$2^{ 18 }$ & 417.428 $\pm$ 1.533 & 22819.912 $\pm$ 569.501 & 0.131 $\pm$ 0.003 \\
\\
$2^{ 19 }$ & 587.52 $\pm$ 2.102 & 44421.93 $\pm$ 1153.896 & 0.129 $\pm$ 0.003 \\
\\
$2^{ 20 }$ & 838.991 $\pm$ 3.097 & 95064.457 $\pm$ 2590.339 & 0.135 $\pm$ 0.003 \\
\\
\hline
\hline
\end{tabular}
\caption{
Mean value and variance of square gyration radius ($R_g^2$) as a function of $N$ and $\beta\mu=0$ for Pr\"ufer-sampled trees of $N$ nodes embedded on the $3d$ FCC lattice.
$\langle R_g^2\rangle$ and ${\rm var}(R_g^2)$ are expressed in [lattice spacing]$^2$ and [lattice spacing]$^4$ units, respectively.
For each $N$, the generated sample is constituted of $10^4$ independent tree configurations.
The last column, the ratio ``${\rm var}(R_g^2) / \langle R_g^2\rangle^2$'', which is associated to the variation in the elastic free energy of the polymer due to a variation of the radius of gyration~\cite{DaoudJoanny1981}, takes a well-defined value $\approx 0.13$ in the large-tree regime.
}
\label{tab:Moments}
\end{table}

\end{document}